\documentclass[10pt]{article}

\usepackage{bm,color,bbm}
\usepackage{ulem} 
\usepackage{hyperref, mathtools,graphicx}

\newcommand{\beq}{\begin{equation}}
\newcommand{\eeq}{\end{equation}}
\newcommand{\bqa}{\begin{eqnarray}}
\newcommand{\eqa}{\end{eqnarray}}

\newcommand{\smallfrac}[2]{\mbox{$\frac{#1}{#2}$}}

\newcommand{\half}{\smallfrac{1}{2}}

\newcommand{\sq}[1]{\left[ {#1} \right]}

\newcommand{\tr}[1]{{\rm Tr}\sq{ {#1} }}

\definecolor{maroon}{rgb}{0.7,0,0}

\definecolor{ngreen}{rgb}{0.3,0.7,0.3}

\definecolor{golden}{rgb}{0.8,0.6,0.1}

\begin{document}
\title{Comment on ``Gleason-Type Theorem for Projective Measurements,
	Including Qubits'' by F. De Zela}
\author{Michael J. W. Hall\\
Centre for Quantum Dynamics,\\ Griffith University, Brisbane, QLD 4111, Australia}
\date{}
\maketitle



\begin{abstract}
It has recently been claimed by De Zela that Gleason's theorem, for probability measures on the lattice of projection operators, can be extended to qubits by adding assumptions related to continuity and the existence of `eigenstates'. This amounts to a claim of the derivation of Born's rule for Hermitian qubit observables. I point out a simple counterexample, and the flaw in De Zela's derivation (these are equally applicable to the repetition of the derivation given in a recent Reply). I also briefly discuss a valid extension to qubits by Busch.
\end{abstract}


\section{Introduction}

Gleason's theorem is an important result in axiomatic quantum mechanics~\cite{gleason}, providing a derivation of Born's rule for quantum probabilities within the context of quantum logic. However, it is only applicable to quantum systems with Hilbert spaces of at least three dimensions, as may easily be demonstrated via counterexamples. Busch has successfully extended Gleason's theorem to qubits~\cite{busch}, albeit at the cost of a strong additional assumption.

Recently, De Zela has claimed a different extension to qubits, based on quite weak additional assumptions related to continuity and the existence of `eigenstates'~\cite{dezela}.  Unfortunately, De Zela's derivation is flawed, as will be shown here by identifying the flaw explicitly and giving a simple counterexample.  While the latter is somewhat trivial, it seems that now there is a published paper~\cite{dezela}, with several (uncritical) citations thereto~\cite{a}, there is value in discussing the relevant issues in a Comment. 

The concepts underlying Gleason's theorem are briefly reviewed in section~2, and its failure for qubits explored in section~3. The interpretation of the strong assumption used by Busch to extend the theorem to qubits is discussed in section~4, and the failure of De Zela's attempted extension in section~5. 

\section{Gleason's theorem and Born's rule}

Gleason's theorem is motivated by the assumption that experimental yes/no propositions correspond to the lattice of projections onto closed subspaces of a separable Hilbert space~\cite{birk}. For projections $P$ and $Q$, the logical operations $P\land Q$, $P\lor Q$ and $\lnot P$ correspond, respectively, to the intersection, linear span and orthogonal complement of the associated subspaces. In particular, one has
\beq
\lnot P=\bm 1-P, \qquad P\land\lnot P = \bm 0,\qquad P\lor\lnot P = \bm 1,
\eeq
where $\bm 0$ is the zero operator, corresponding to the trivially always-false proposition (projecting onto the trivial subspace $\{0\}$), and $\bm 1$ is the unit operator, corresponding to the trivially always-true proposition (projecting onto the whole Hilbert space). Note that $\lnot \bm 0= \bm 1$. 

Under these logical operations the set of projections forms an orthomodular lattice, rather than a Boolean lattice as would be formed by a set of classical yes/no propositions~\cite{birk,belt}. In particular, the distributive law $P\land(Q\lor R)=(P\land Q)\lor (P\land R)$ for Boolean lattices does not hold for general projections $P$, $Q$ and $R$.

However, significantly, any set of mutually orthogonal projections $P_1,P_2,\dots $, corresponding to a set of mutually orthogonal subspaces, does generate a Boolean lattice under the above logical operations. It is easy to show that the elements of such a set commute, and that
\beq \label{xor}
P_j\lor P_k=P_j+ P_k,\qquad P_j\land P_k= P_jP_k = \bm 0, \qquad {\rm for~~}j\neq k .
\eeq
Hence,  it is natural to interpret mutually orthogonal projectors as representing disjoint outcomes of a single experiment, and the Boolean lattice generated by these projectors as a classical logical structure for the outcomes of such an experiment~\cite{birk,belt,hall}. 

Now, for any given experiment the probabilities of disjoint experimental outcomes must be additive, i.e, 
\[ p(A=a \lor A=b)=p(A=a)+p(A=b) {\rm ~~~for ~~~}a\neq b . \]
This motivates the main assumption required for Gleason's theorem~\cite{gleason}. In particular, for some preparation procedure $s$, let $ p_s(P)\in [0,1]$ denote the probability that the experimental proposition corresponding to projection $P$ will be verified. Hence, noting the first equality in Eq.~(\ref{xor}), the above interpretation of mutually orthogonal projections requires that 
\beq \label{sum}
p_s(P_1+P_2+\dots) = p_s(P_1) + p_s(P_2) + \dots 
\eeq
for any set of mutually orthogonal projectors $P_1,P_2,\dots$. One must further have
\beq \label{one}
p_s(\bm 1) = 1,
\eeq
corresponding to verification of the trivially-true proposition with probability one.

Gleason's theorem states that {\it if the Hilbert space is at least 3-dimensional, and the assumptions in Eqs.~(\ref{sum}) and~(\ref{one}) hold, then there is a density operator $\rho_s$ on the Hilbert space such that}~\cite{gleason}
\beq \label{thm}
 p_s(P) = \tr{\rho_s P}.
\eeq
For the case of a one-dimensional projection, $P_\psi=|\psi\rangle\langle\psi|$, and a pure state,  $\rho_\phi=|\phi\rangle\langle\phi|$ (i.e., an extreme point of the convex set of density operators), this theorem reduces to Born's rule:
\beq
p_\phi(P_\psi) = |\langle \psi|\phi\rangle|^2 .
\eeq
Thus, in the quantum logic approach, Born's rule follows once one has axioms sufficient to identify experimental propositions with Hilbert space projections. 

It is of interest to remark that Bell adapted Gleason's proof to show that one cannot consistently assign pre-existing definite outcomes of quantum measurements to the set of projections, for Hilbert spaces of three or more dimensions~\cite{bell}. This is an example of quantum contextuality, analogous to the Kochen-Specker theorem~\cite{ks}, but independent  of the latter and requiring consideration of a continuum of projections rather than a finite number.

\section{The problem with qubits}

Gleason's theorem in Eq.~(\ref{thm}) only applies to Hilbert spaces of three or more dimensions. For qubits it is simple to find counterexamples.

For example, note for a qubit Hilbert space that any one-dimensional projection $P_\psi$, and its orthogonal complement $\lnot P_\psi$, have the respective forms
\beq
P_\psi = |\psi\rangle\langle\psi| = \half(\bm 1+\sigma\cdot n^\psi),\qquad \lnot P_\psi = \bm 1-|\psi\rangle\langle\psi| = \half(\bm 1-\sigma\cdot n^\psi) ,
\eeq
where $\sigma=(\sigma_x,\sigma_y,\sigma_z)$ is the vector of Pauli sigma matrices and $n^\psi:=\langle \psi|\sigma|\psi\rangle$ denotes the unit Bloch vector corresponding to $P_\psi$. Now define the function $p_{s}(P)$ on the lattice of qubit projections by
\beq \label{ex}
p_{s}(\bm 0):=0,\quad p_{s}(\bm 1):=1,\quad p_{s}(P_\psi):= \half\left\{1 + \left(\tr{P_\psi \sigma_z}\right)^3\right\} =\half\left[1+(n^\psi_z)^3\right] .
\eeq
Clearly, this function is nonlinear in $P_\psi$, and hence cannot be generated by a density operator as per Eq.~(\ref{thm}).  However, using $\tr{\sigma_z}=0$, one has
\[ p_{s}(\lnot P_\psi) = \half\left\{1 + \left(\tr{(\bm 1-P_\psi) \sigma_z}\right)^3\right\} =\half\left\{1 - \left(\tr{P_\psi \sigma_z}\right)^3\right\} = 1 - p_{s}(P_\psi) .\]
Hence, both Eqs.~(\ref{sum}) and~(\ref{one}) are satisfied.

It is useful for later purposes to note that this counterexample for qubits may be generalised further, to the probability functions
\beq  \label{genex}
p_m(\bm 0):=0,\quad p_m(\bm 1):=1,\quad p_m(P_\psi) := \half\left[ 1 + f(m\cdot n^\psi)\right] ,
\eeq
where $m$ denotes any unit 3-vector, and $f(x)$ is any real nonlinear function from  the interval $[-1,1]$ into itself satisfying $f(-x)=-f(x)$ and $f(1)=1$. Eq.~(\ref{ex}) corresponds to $m=(0,0,1)$ and $f(x)=x^3$.

\section{Extending Gleason's theorem to qubits}

To generalise Gleason's theorem to qubits, it follows from the previous section that one must make at least one further assumption in addition to Eqs.~(\ref{sum}) and~(\ref{one}). For example, Busch requires that Eq.~(\ref{sum}) be strengthened to~\cite{busch}
\beq \label{busch}
p_s(E_1+E_2+\dots) = p_s(E_1) + p_s(E_2) + \dots
\eeq
for any set of  operators $\{ E_j\}$ satisfying $E_j\geq \bm 0 $ and $E_1+E_2+\dots\leq \bm 1$. 
This assumption, together with Eq.~(\ref{one}), leads straightforwardly to the result that $p_s$ must have a density operator representation as per Eq.~(\ref{thm}), even for qubit Hilbert spaces~\cite{busch}. It should, however, be remarked that the  motivation for the above strengthened assumption is somewhat weaker than the ``logical'' motivation for Gleason's original theorem.

In particular, a set of operators as per the above assumption corresponds to a subset of a positive-operator-valued measure (POVM). However, while in quantum mechanics one may associate the elements of a POVM with disjoint outcomes of a physical experiment, such elements do not generate a Boolean sublattice of operators. Hence, a different motivation for Eq.~(\ref{busch}) is required.

One must be careful to avoid circularity in motivating Eq.~(\ref{busch}). The {\it a priori} appearance of `probability operators' (POVM elements), with $\bm 0\leq E_j \leq \bm 1$, is itself difficult in this regard. For example, within standard quantum mechanics, they appear in two natural ways.  First, they may be regarded as a formal extension of Born's rule: if one assumes that probabilities are of the form $\langle\psi |A|\psi\rangle$ for some operator $A$, then $A$ must be a probability operator, i.e., $\bm 0\leq A\leq \bm 1$.  However, this assumes from the start that $p_s$ is linear with respect to $|\psi\rangle\langle\psi|$, which is tantamount to assuming it is of the desired form in Eq.~(\ref{thm}). 

Second,  any probability operator may be regarded as corresponding to a yes/no proposition associated with the joint measurement of a projection on a system plus ancilla (note that, within standard quantum mechanics, this already requires the ancilla to be described by a density operator). However, as soon as one considers the set of projections on the joint Hilbert space of a qubit and some ancilla, then the relevant Hilbert space dimension becomes greater than two (one is no longer dealing with a qubit!), and one may simply revert to the original derivation of Gleason's theorem.

Note further that while it is natural to consider a yes/no proposition $E$ that corresponds to testing one of a set of projections $P, Q,R,\dots$ with respective probabilities $p, q, r,\dots$ (satisfying $p+q+r\dots = 1$), yielding the consistency requirement
\beq
p_s(E) = p\,p_s(P) + q\,p_s(Q) +r\,p_s(R) +\dots,
\eeq
there is no {\it a priori} reason to demand that $E$ corresponds to the probability operator $pP+qQ+rR+\dots$ (indeed this would immediately imply the linearity of $p_s$, and hence a density operator representation, even for qubits).

In any case, once one does decide to consider the set of probability operators, for whatever reason, then the strengthened assumption in Eq.~(\ref{busch}) must itself still be motivated. For example, one can postulate that if two probability operators $E$, $F$ satisfy $E+F\leq \bm 1$, then they must correspond to disjoint outcomes of some physical experiment. Note that a similar postulate is implicit to the weaker assumption in Eq.~(\ref{sum})~\cite{hall}, although in the latter case one has a ``logical'' motivation for considering the set of mutually orthogonal projections. 

Putting the interesting question of motivation aside, Eq.~(\ref{busch}) most certainly provides a suitable assumption for extending Gleason's theorem to the qubit case~\cite{busch}.  In contrast, the additional assumptions proposed recently by De Zela~\cite{dezela} do not, as will now be discussed.

\section{Why De Zela's extension fails}

In contrast to Busch's approach, De Zela remains within the ambit of quantum logic concepts. In particular, attention is restricted to  the set of projections on Hilbert space, rather than broadened to the set of all probability operators~\cite{dezela}. However, he proposes, in addition to Eqs.~(\ref{sum}) and~(\ref{one}), that for each rank-1 projection $P_\phi=|\phi\rangle\langle\phi|$ on the Hilbert space there is an associated probability measure $p_\phi$ such that
\begin{itemize} \item[\bf (i)] $p_\phi(P_\psi)$ is continuous with respect to the parameters used to specify the set of rank-1 projections $\{ P_\psi\}$, and
	\item[\bf (ii)] $p_\phi(P_\phi) = 1$.
\end{itemize}
Note that these are physically reasonable assumptions, with the first corresponding to the notion that similar projections have similar probabilities, and the second to the existence of an `eigenstate' for each rank-1 projection~\cite{gisin}.

However, despite the claim made by De Zela \cite{dezela}, these additional assumptions are not sufficient to derive Gleason's theorem for qubits. In particular, for a qubit  with rank-1 projection $P_\phi$ and corresponding unit Bloch vector $n^\phi$, consider a probability measure as per Eq.~(\ref{genex}) with $m=n^\phi$, i.e.,
\beq \label{counter}
p_{\phi}(P_\psi) := \half\left[1 + f(n^\phi\cdot n^\psi)\right] = \half\left\{ 1 + f(2\,\tr{P_\phi P_\psi}-1)\right\} ,
\eeq
where $f(x)$ is any nonlinear function mapping the interval $[-1,1]$ into itself, with $f(-x)=-f(x)$ and $f(1)=1$. Note that, as well as satisfying Eqs.~(\ref{sum}) and~(\ref{one}), this measure satisfies both of De Zela's above additional assumptions~(i) and~(ii), provided that $f(x)$ is restricted to be a continuous function---e.g., $f(x)=x^3$ as in Eq.~(\ref{ex}).  Hence, since nonlinearity of $p_\phi$ with respect to $P_\psi$  implies that it cannot be generated by some density operator as per Eq.~(\ref{thm}), it provides a counterexample to De Zela's claim.

So, where does De Zela's derivation of Gleason's theorem for qubits fail? It is in the misuse of a theorem of Gudder: any  continuous mapping of a vector to the real numbers, $g(v)$, that is also orthogonally additive, i.e., $g(u+v)=g(u)+g(v)$ for $u\cdot v=0$, must have the form $g(v)= a\,v \cdot v+ b\cdot v$ for some constant $a$ and fixed vector $b$~\cite{gudder}. 
De Zela applies Gudder's theorem to orthogonally-additive continuous functions of 4-vectors, and only afterwards considers the two-dimensional submanifold $\{v=(1,n^\psi)\}$.  However, assumption (i) above of De Zela does not require the continuity of $p_\phi(P_\psi)$ with respect to general 4-vectors, but only with respect to the unit 3-vectors $n^\psi$ that parameterise the projections $P_\psi$.  Hence, Gudder's theorem simply does not apply---as further evidenced, of course, by the above explicit counterexample to De Zela's claim.

\section{Remarks}

There is renewed  interest in attempting to derive as much of the formalism of quantum mechanics as possible from a starting set of axioms. In my opinion, given the existence of the quantum measurement problem and the difficulty in unifying quantum mechanics with gravity, there should perhaps be even more interest in attempting to consistently tweak quantum mechanics without breaking it (e.g., without introducing observable nonlocal effects) \cite{miw,hr}. However, successful extensions of Gleason's theorem to qubits, such as given by Busch~\cite{busch}, particularly if based on strong physical motivations, are worth pursuing.


\end{document}